\documentclass[aip,jcp,reprint,amsmath,amssymb]{revtex4-1}
\usepackage{graphicx}
\usepackage{dcolumn}
\usepackage{bm}
\usepackage[utf8]{inputenc}
\usepackage[T1]{fontenc}
\usepackage{mathptmx}
\usepackage{etoolbox,float}
\usepackage{booktabs}

\makeatletter
\def\@email#1#2{%
 \endgroup
 \patchcmd{\titleblock@produce}
  {\frontmatter@RRAPformat}
  {\frontmatter@RRAPformat{\produce@RRAP{*#1\href{mailto:#2}{#2}}}\frontmatter@RRAPformat}
  {}{}
}%
\makeatother

\begin{document}
\preprint{}

\title{Enhancing the reactivity of Si(100)-Cl toward PBr$_3$ by charging Si dangling bonds}

\author{T. V. Pavlova}
 \email{pavlova@kapella.gpi.ru}
\affiliation{Prokhorov General Physics Institute of the Russian Academy of Sciences, Vavilov str. 38, 119991 Moscow, Russia}
\affiliation{HSE University, Myasnitskaya str. 20, 101000 Moscow, Russia}

\author{V. M. Shevlyuga}
\affiliation{Prokhorov General Physics Institute of the Russian Academy of Sciences, Vavilov str. 38, 119991 Moscow, Russia}

\date{\today}

\begin{abstract}

The interaction of the PBr$_3$ molecule with Si dangling bonds (DBs) on a chlorinated Si(100) surface was studied. The DBs were charged in a scanning tunneling microscope (STM) and then exposed to PBr$_3$ directly in the STM chamber. Uncharged DBs rarely react with molecules. On the contrary, almost all positively charged DBs were filled with molecule fragments. As a result of the PBr$_3$ interaction with the positively charged DB, the molecule dissociated into PBr$_2$ and Br with the formation of a Si-Br bond and PBr$_2$ desorption. These findings show that charged DBs significantly modify the reactivity of the surface towards PBr$_3$. Additionally, we calculated PH$_3$ adsorption on a Si(100)-2$\times$1-H surface with DBs and found that the DB charge also has a significant impact. As a result, we demonstrated that the positively charged DB with a doubly unoccupied state enhances the adsorption of molecules with a lone pair of electrons.

\end{abstract}

\maketitle

\section{Introduction}
Silicon dangling bonds (DBs) appear on a Si(100)-2$\times$1 surface covered with an adsorbate monolayer if one adatom is absent. These DBs act as active sites for molecules that include phosphine \cite{2003Schofield}, arsine \cite{2020Stock}, hydrocarbons \cite{2011Zikovsky}, HCl \cite{2011Li}, and I$_2$ \cite{2012Ferng}, altering the local reactivity of the surface. The DB can exist in three charge states, holding zero (DB$^+$), one (DB$^0$), or two (DB$^-$) electrons \cite{2013Schofield}. Depending on the charge state, DBs alter the reactivity of the surface in different ways, forming various local charges on it. For example, as was shown for BF$_3$ adsorption on negatively charged DB \cite{2002Cao}, nucleophilic DBs$^-$ promote the adsorption of molecules having an empty lone pair of electrons. Furthermore, it was demonstrated that the DB charge influences the self-assembly of organic nanostructures \cite{2012Ryan, 2014Piva}. By removing adatoms from specific sites, single DBs can be precisely created in a scanning tunneling microscope (STM) \cite{1995Shen, 2017Moller, 2018Achal, 2018Randall}, and the charge of individual DBs can also be changed in an STM \cite{2013Bellec, 2015Labidi}. Therefore, by charging different DBs differently, it is possible to control the selectivity of the reaction on the surface.

The PH$_3$ reaction with DBs on a Si(100) surface covered with a hydrogen monolayer is used for the exact positioning of dopant atoms on the surface \cite{2001OBrien, 2003Schofield, 2019He}. For this purpose, it is possible to use DBs on halogenated surfaces  \cite{2018Pavlova, 2019Dwyer, 2020Pavlova, 2022Pavlova} and other molecules such as BCl$_3$ \cite{2020Silva-Quinones, 2021DwyerACS} and AlCl$_3$ \cite {2021Radue}. Molecules such as PH$_3$, PCl$_3$, and PBr$_3$ are trigonal pyramidal in shape and have a lone pair of electrons. We can expect that the adsorption of such molecules will be facilitated by the DB$^+$, which has a doubly unoccupied state. And, accordingly, the DB$^-$ with two occupied states will make more favorable the adsorption of a molecule with a lone pair of electrons, such as BCl$_3$ and BF$_3$ \cite{2002Cao}. Note that charging of the DBs on the halogenated Si(100) surface has already been demonstrated \cite{2022Pavlova}.

In this work, the PBr$_3$ adsorption on neutral and positively charged DBs on the Si(100)-2$\times$1-Cl surface was studied. In an STM, we charged the DBs and then performed PBr$_3$ adsorption. We found significant differences in the PBr$_3$ adsorption on DBs depending on the charge. To support our results, we performed DFT calculations of the PBr$_3$ adsorption on charged DBs. We also calculated PH$_3$ adsorption onto charged DBs on a hydrogenated Si(100) surface and found a strong influence of the DB charge.

\begin{figure*}[t!]
 \begin{center}
 \includegraphics[width=\linewidth]{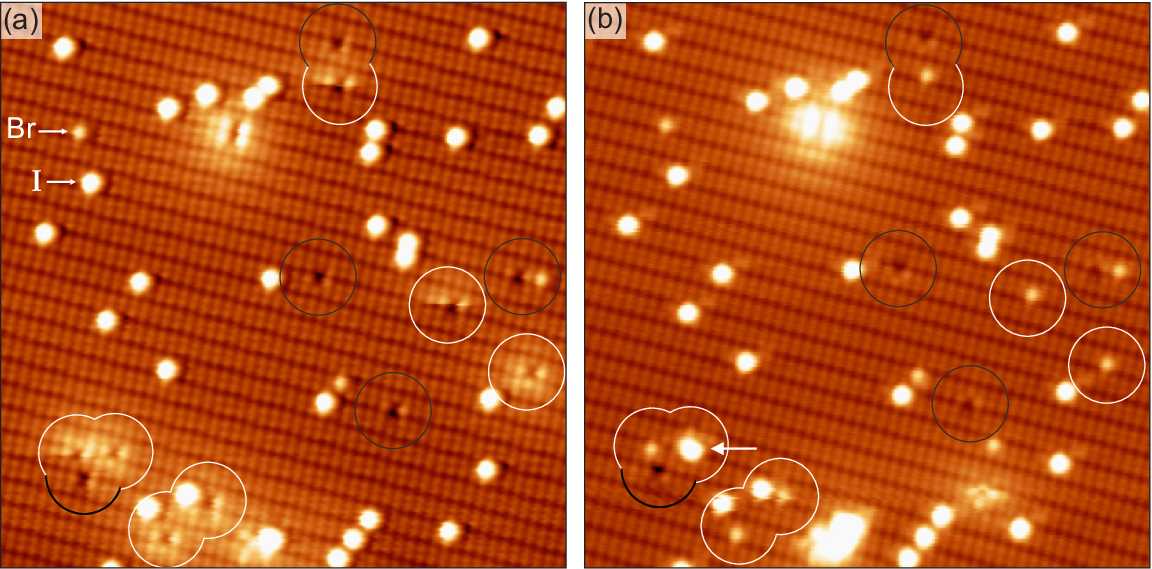}
\caption{\label{fig1} Empty state STM images (18.1$\times$18.1\,nm$^2$,  I$_t$ = 2.0\,nA) of the (a) initial Si(100)-2$\times$1 surface covered with an incomplete Cl monolayer ($U_s =3.1$\,V) and (b) the same surface area after exposure to PBr$_3$ ($U_s =2.5$\,V). Vacancies with the DB$^0$ and DB$^+$ are indicated with black and white circles, respectively. Bright spots are I and Br atoms. After PBr$_3$ adsorption, the DBs$^+$ are mostly filled with Br atoms, while the DBs$^0$ are empty. The arrow marks an object in the DBs$^+$, which may be some fragment of the PBr$_3$ molecule. The slow scan direction proceeded from bottom to top.}
\end{center}
\end{figure*}

\section{Experimental and computational details}

The investigations were carried out in an ultra-high vacuum setup with a base pressure of 5$\times$10$^{-11}$\,Torr equipped with the STM (GPI CRYO, SigmaScan Ltd.) operating at 77\,K. We used the B-doped Si(100) samples (1\,$\Omega$\,cm). The wafers were outgassed at 870\,K for several days in UHV, and then flash-annealed at 1470\,K. The sample temperature was measured using an optical pyrometer. Note that as a result of this annealing, the near-surface region was depleted of impurities \cite{2012Pitters}. To prevent water adsorption, Cl$_2$ was introduced at a sample temperature of 370--420\,K after the flash heating was switched off. The sample temperature during Cl$_2$ deposition was monitored by thermocouple mounted to the sample holder. A fine leak valve delivered molecular chlorine to the Si(100) surface at a partial pressure in the chamber of $2 \cdot 10^{-9}$\,Torr for 2.5 seconds. Due to the presence of iodine and bromine in the inlet line utilized in earlier experiments, they were adsorbed on the Si(100) surface together with chlorine. The amounts of iodine and bromine on the surface were so low that their presence had no effect on the results of the experiment but helped to find the same surface area after adsorption. More specifically, the close proximity of I and Br did not affect the measurements. PBr$_3$ gas was delivered into the STM chamber through a tube that passed via holes in cold screens and located about 10\,cm away from the Si(100) surface. The PBr$_3$ adsorption was carried out with a partial pressure of $3 \cdot 10^{-10}$\,Torr for three minutes at 77\,K. Note that there was no iodine or bromine in the inlet line used for PBr$_3$. To eliminate P atoms from the surface from the earlier experiment, the sample was heated at 1170\,K  for several hours \cite{2005Brown}. Mechanically cut Pt-Ir tips were used. Voltage ($U_s$) was applied to the sample. To process all STM images, the WSXM software \cite{WSXM} was used.

The spin-polarized DFT calculations were carried out in the Vienna \textit{ab initio} simulation package (VASP) \cite{1993Kresse, 1996Kresse} by employing the projector augmented wave approach \cite{1999Kresse, 1994Blochl}. The exchange-correlation potential was treated within the framework of Perdew-Burke-Ernzerhof (PBE) generalized gradient approximation \cite{1996Perdew}. The DFT-D2 approach developed by Grimme was used for van der Waals correction \cite{2006Grimme}. The cutoff of the plane wave energy was set at 400 eV. To model the Si(100)-2$\times$1 surface, a slab consisting of 16 Si layers with periodic 6$\times$6 supercells was used. Chlorine atoms were positioned on the upper surface to form a Si(100)-2$\times$1-Cl structure, whereas the hydrogens covered the bottom surface. Only the two bottom silicon layers were fixed during optimization. The residual forces acting on the relaxed atoms were less than 0.01\,eV/\,{\AA}. To prevent surface-surface interactions, the slabs were separated by a vacuum space of 16\,{\AA}. The reciprocal cell was integrated using a $\Gamma$-centered 4$\times$4$\times$1 k-point mesh. To simulate positive and negative charge states of DBs, the electron was removed or added to the supercell, respectively. The adsorption energies of PBr$_3$ were calculated as the difference between the total energy of the surface with the molecule and the total energies of the surface and PBr$_3$ in the gaseous phase.

\section{Results and discussion}

\subsection{Experimental Results}

Figure~\ref{fig1}a shows the Si(100)-2$\times$1 surface with Cl vacancies containing neutral and positively charged DBs.
Initially, the surface contained only neutral DBs. To charge the DBs, we used the procedure described in Ref.~\cite{2022Pavlova}. Charging of the DBs occurs due to the strong tip induced band bending (TIBB). When a positive voltage is applied to the sample, the electron of the DB moves away from the negative local potential induced by the tip \cite{2008Teichmann}. To charge several DBs, we gradually increased the scan voltage from about 2 V to 3 V. We were able to charge only a part of the DBs while leaving the remainder uncharged since DBs have different charging voltages due to surrounding impurities and defects \cite{2015Labidi, 2022Pavlova}. The charge state of the DB is easily distinguished in an STM image because positively charged DBs have a bright halo, whereas neutral ones do not \cite{2022Pavlova}. If charging occurs when the tip passes over a dangling bond, the removal of an electron is visualized as a break through the scan in the STM image. As shown in Fig.~\ref{fig1}a, some neutral DBs become positively charged after the tip passes through them, and a bright halo appears.

Before PBr$_3$ adsorption, we retracted the tip from the surface. When the tip is removed, the DB maintains its charge state \cite{2022Pavlova}. To prove this, we charged some of the DBs on the surface, then retracted the tip, and when we scanned the same area after several hours, we found that the DBs kept their charge. We did not use negatively charged DBs since they have a doubly occupied state and hence are inefficient for adsorption of molecules with a lone pair of electrons, such as PBr$_3$. However, they may be relevant for molecules having an empty lone pair of electrons, such as BCl$_3$, BBr$_3$, and BF$_3$.

\begin{figure}[h]
 \includegraphics[width=\linewidth]{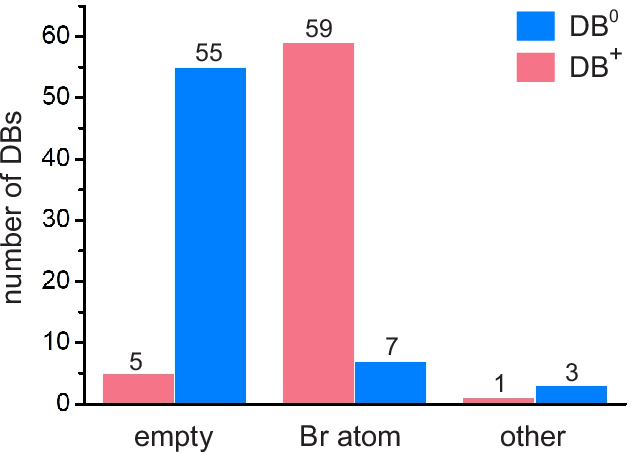}
\caption{\label{fig2} Histogram illustrating the filling of dangling bonds after PBr$_3$ adsorption on a Si(100)-2$\times$1 surface covered with an incomplete Cl monolayer (based on the analysis of the STM images with 65 DBs$^0$ and 65 DBs$^+$).}
\end{figure}

After PBr$_3$ adsorption, all neutral DBs remained empty, whereas bright objects appeared in the positively charged ones (Fig.~\ref{fig1}b). Such a bright object looks like a bromine atom in Fig.~\ref{fig1}a. In rare cases, other objects appeared in the DBs. Note that the STM images before and after adsorption were recorded at different tip states, which were slightly revealed by the presence of a small depression on the right side of I and Br atoms before (Fig.~\ref{fig1}a) and a very faint protrusion after adsorption (Fig.~\ref{fig1}b). Figure~\ref{fig2} shows a diagram of the distribution of the adsorbate presence in the neutral and positively charged DBs after the surface exposure to PBr$_3$. In neutral DBs, nothing was observed in 91$\%$ of cases, while in positively charged DBs, bromine was observed in 85$\%$ of cases.

\subsection{Computational Results}

\begin{figure}[h]
 \includegraphics[width=\linewidth]{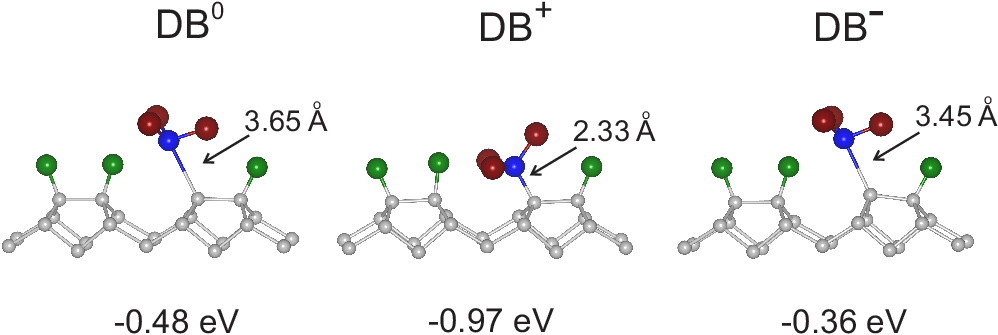}
\caption{\label{fig3} Optimized structures after PBr$_3$ adsorption by a phosphorus atom down on the Si(100)-2$\times$1 surface with DBs in different charge states. Si atoms are shown in gray, Cl in green, Br in red, and P in blue.}
\end{figure}

We studied the PBr$_3$ adsorption onto charged DBs on the Si(100)-2$\times$1 surface in two scenarios: when the molecule is approached by a phosphorus atom and a bromine atom. Figure~\ref{fig3} shows the optimized structures for the molecule with the P atom bonded to the Si atom. Only when PBr$_3$ adsorbs on the DB$^+$, the P-Si bond is sufficiently short, indicating a strong interaction between the P and Si atoms. For the DB$^0$ and DB$^+$, the P-Si bond length is about 3.5\,{\AA}, demonstrating a weak interaction between the molecule and the surface. Adsorption energy calculations confirm this conclusion, since the PBr$_3$ adsorption on the DB$^+$ is twice as favorable as on the DB$^0$. Thus, as expected, the PBr$_3$ molecule with a lone pair of electrons is more favorably adsorbed on the DB$^+$ with a doubly unoccupied state.

In the experiment, however, we identified a Br atom in the majority of DBs$^+$ rather than the PBr$_3$ molecule. According to the calculations, when a bond is formed between the Br atom of the PBr$_3$ molecule and the Si atom containing the DB$^0$, DB$^+$, or DB$^-$, the PBr$_3$ molecule dissociates into Br and PBr$_2$. The PBr$_2$ fragment is removed into the vacuum since we did not find similar fragments with a concentration close to the bromine concentration in the STM images. The adsorption energy of Br in the DB$^0$ with PBr$_2$ in the gas phase is $-1.14$\,eV, which is significantly more favorable than the adsorption of PBr$_3$ in the DB$^0$ ($-0.48$\,eV). When PBr$_3$ is adsorbed onto charged DBs and the Si-Br bond is formed, the charge is removed from the DB. The energies of PBr$_3$ adsorption on the DB$^+$ and DB$^-$ by a bromine atom down were not calculated because the result depended on where the residual charge from the DB moves. In the experiment, an electron (hole) can be captured by other defects, impurities, or other DBs. We did not simulate this situation because modeling electron capture by defects is challenging and there is no experimental evidence of where the charge is relocated.

\begin{figure}[h!]
 \includegraphics[width=\linewidth]{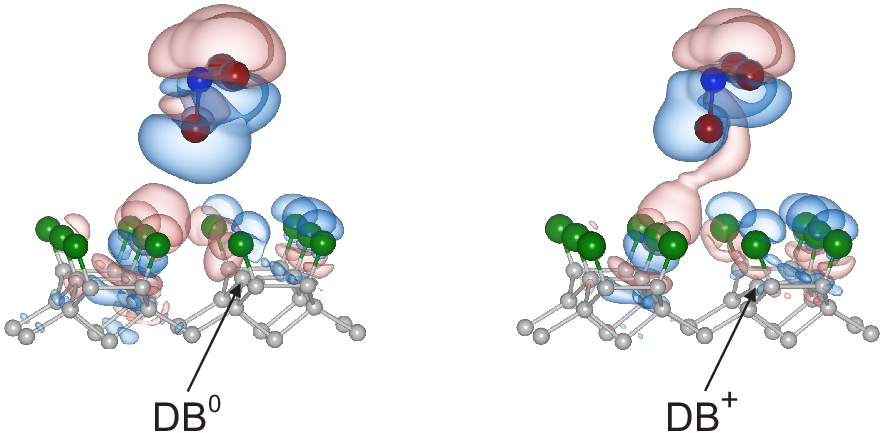}
\caption{\label{fig4} Charge density difference plots for the PBr$_3$ molecule in 5\,{\AA} above the Si(100)-2$\times$1 surface with the DB$^0$ and DB$^+$. The isosurface level is taken as $4 \cdot 10^{-6}$e/\,{\AA}$^3$. Red regions denote the accumulation of charge density, while blue regions denote charge density depletion. To avoid the molecule approaching the surface, the coordinates of the PBr$_3$ molecule and the surface were first optimized separately, and the electronic density was then optimized for the entire system (the PBr$_3$ molecule and the surface). Si atoms are shown in gray, Cl in green, Br in red, and P in blue.}
\end{figure}

We can understand the more favorable PBr$_3$ adsorption on the DB$^+$ than on the DB$^0$ by considering the electron density distribution in both cases. Figure~\ref{fig4} shows the electron density distribution when PBr$_3$ is fixed at a distance of 5\,{\AA} from the bottom Br to the plane of the Cl monolayer. The PBr$_3$ molecule has a common electron density with the surface with the DB$^+$ in contrast to the surface with the DB$^0$. According to the electron density distribution (Fig.~\ref{fig4}), PBr$_3$ interacts with the surface with the DB$^+$ at a greater distance than the DB$^0$. This may explain why bromine is found in the DB$^+$ with a significantly greater probability than in the DB$^0$.

Interestingly, when PBr$_3$ approaches the surface with the DB$^+$, the P atom interacts with the Cl atom located on the neighboring dimer with the DB$^+$ (Fig.~\ref{fig4}). This result was unexpected since the molecule was anticipated to interact directly with the electron density of the DB$^+$. Unlike Cl atoms farther away from the DB$^+$, the Cl atom on the dimer adjacent to the DB$^+$ has a partial electron density hybridized with the DB$^+$. Indeed, in the empty state STM image, a bright halo caused by the unoccupied states of the DB$^+$ extends to the Cl atoms next to the DB$^+$ \cite{2022Pavlova}. Thus, our calculations suggest the following scenario for the PBr$_3$ interaction with the surface with the DB$^+$. After the P atom is attracted to the Cl atom, the Br atom begins to interact with the DB$^+$, forming a Si-Br bond, and then the molecule dissociates, with PBr$_2$ moving away from the surface.

Additionally, we calculated the phosphine adsorption on the Si(100)-2$\times$1-H surfaces with charged DBs. The phosphine molecule, like PBr$_3$, has a lone pair of electrons and thereby interacts much more strongly with positively charged bonds. In the case of adsorption of the entire molecule into a vacancy with the formation of a P-Si bond, the PH$_3$ adsorption energy on the DB$^+$ was found to be $-1.2$\,eV, which is $0.8$\,eV more advantageous than adsorption on the DB$^0$. However, there is a significant difference in PH$_3$ and PBr$_3$ adsorption when a molecule approaches a surface with an atom other than phosphorus. When PH$_3$ is approached by the H atom, it does not dissociate on the DB with any charge, since the final state with the H atom on the DB and PH$_2$ in a vacuum is less favorable than the initial state with PH$_3$ above the surface. Thus, the DBs will not be capped by hydrogen as a result of PH$_3$ adsorption, unlike PBr$_3$ adsorption, which is most favored by dissociation and capping of the DBs by bromine.

\section{Conclusions}

We studied the effect of the DBs charge on the Si(100)-2$\times$1-Cl surface reactivity toward PBr$_3$. The DBs were charged in an STM, and then PBr$_3$ gas was supplied to them in the same STM chamber. We were, therefore, able to determine the active sites of the reaction depending on the charge of the DB. According to our STM data, the DBs$^+$ are filled with Br atoms, whereas the DBs$^0$ are mostly left empty. If the PBr$_3$ molecule approaches the surface with the Br atom down, it dissociates into PBr$_2$ and Br in the DB, with Br bonding to Si and PBr$_2$ being ejected into vacuum. Our calculations suggest that in this case, first the P atom with a lone pair of electrons interacts with the Cl atom, and then Br bonds with the Si atom. The molecule interacts with the DB$^+$ at a higher distance from the surface than with the DB$^0$, according to electron density analysis, which may explain the higher probability of the DB$^+$ being filled with Br. Adsorption of the PBr$_3$ molecule is more favorable when Br is located closer to the surface than P. Nevertheless, if the P atom is closer to the surface, PBr$_3$ adsorption on the DB$^+$ is significantly more advantageous than on the DB$^0$ or DB$^-$. Our calculations imply that phosphine adsorption to charged DBs on Si(100)-2$\times$1-H will exhibit a similar trend, namely, much more favorable adsorption on the DB$^+$. These results demonstrate that charging the DBs can significantly alter the reactivity of the Si(100)-2$\times$1 surface with an adsorbate monolayer toward molecules like PBr$_3$ and PH$_3$.

\section*{Supplementary Material}
See supplementary material for STM images showing the coadsorption of Cl and Br on Si(100) and visualization of the DB charging on the STM image.

\begin{acknowledgments}
This study was supported by the Russian Science Foundation under grant No. 21-12-00299. We also thank the Joint Supercomputer Center of RAS for providing the computing power.
\end{acknowledgments}

\bibliography{arxiv_Br_in_DB_JCP}
\end{document}